\documentclass[aps,prd,amsmath,amssymb,showpacs,showkeys,nofootinbib,floatfix]{revtex4}
\usepackage{graphicx}% Include figure files
\usepackage{dcolumn}% Align table columns on decimal point
\usepackage{bm}% bold math 
\usepackage{epsfig}
\begin{document}
\title{Effective continuum threshold for vacuum-to-bound-state correlators}
%HEPHY-PUB 845/07
\author{Wolfgang Lucha$^{a}$, Dmitri Melikhov$^{a,b,c}$, Hagop Sazdjian$^{d}$, and Silvano Simula$^{e}$} 
\affiliation{
$^a$HEPHY, %Institute for High Energy Physics,
Austrian Academy of Sciences, Nikolsdorfergasse 18, A-1050 Vienna, Austria\\
$^b$Faculty of Physics, University of Vienna, Boltzmanngasse 5, A-1090 Vienna, Austria\\
$^c$SINP, Moscow State University, 119991 Moscow, Russia\\
$^d$IPN, 
CNRS/IN2P3, Universit\'e Paris-Sud 11, F-91406 Orsay, France\\
$^e$INFN, Sezione di Roma III, Via della Vasca Navale 84, I-00146 Roma, Italy}
\date{\today}
\begin{abstract}
We formulate a new algorithm for obtaining the effective continuum threshold in vacuum-to-bound-state correlators  
--- the basic objects for the calculation of hadron form factors in the method of light-cone sum rules in QCD. 
The effective continuum threshold is one of the key ingredients of the method which to a large extent 
determines the value of the form factor extracted from the relevant correlator. 
In a quantum-mechanical potential model, where the exact form factor is known, the application of our 
algorithm is shown to lead to a dramatic increase of the accuracy of the extracted form factor 
compared to the standard procedures adopted in the method of sum rules in QCD. 
Moreover, the application of our algorithm considerably enlarges the range of the momentum transfer  
where the form factor may be extracted from the correlator. 
\pacs{12.38-t, 11.10.St, 11.55.Hx}
\keywords{Hadron physics, strong interactions, bound states, dispersive sum rules}
\end{abstract}

\maketitle
\section{Introduction}
The extraction of the ground-state parameters from a correlator of quark currents is a cumbersome procedure: 
even if several terms of the expansion for the correlator (i.e., 
the OPE in the case of vacuum-to-vacuum correlators and the twist expansion in the case of  
vacuum-to-hadron correlators) are known precisely, the numerical procedures of the method of 
sum rules \cite{svz,ioffe} cannot determine the true exact value of the bound-state parameter. 
Instead, the method should provide the band of values such that the true hadron parameter has a 
flat probability distribution within this band \cite{svz}. This band is a measure of the systematic, or intrinsic, 
sum-rule uncertainty. 
 
The method of sum rules in QCD contains a set of prescriptions (see e.g.~\cite{ck}) 
which are believed to provide such a systematic error. In QCD this, however, always remains a conjecture ---  
it is impossible to prove that and even to check unambiguously whether the range provided by the standard sum-rule  
procedures indeed contains the actual value of the bound-state parameter. 

The only possibility to check the reliability of the corresponding procedures is to apply the method to a 
problem where the parameters of the ground state may be calculated independently and exactly. 
Presently, only quantum-mechanical potential models provide such a possibility. 
(For a discussion of many aspects of sum rules in quantum mechanics we refer to 
\cite{nsvz,nsvz1,qmsr,orsay,ms_inclusive,radyushkin2001,bakulev}).

A simple harmonic-oscillator (HO) potential model is a perfect tool to achieve this goal: 
it possesses two essential features of QCD --- 
confinement and asymptotic freedom \cite{nsvz} --- and has the following advantages: 
(i) the bound-state parameters (masses, wave functions, form factors) are known precisely;
(ii) direct analogues of the QCD correlators may be calculated exactly. 

Making use of this model, we have already studied the extraction of ground-state parameters from 
different types of correlators: namely, the ground-state decay 
constant from a two-point vacuum-to-vacuum correlator \cite{lms_2ptsr}, the form factor from a three-point 
vacuum-to-vacuum correlator 
\cite{lms_3ptsr}, and the form factor from a vacuum-to-hadron correlator \cite{m_lcsr}. We have demonstrated 
that the standard procedures adopted in the literature for obtaining the systematic errors do not work properly: 
for all types of correlators the true known value of the bound-state parameter was shown to lie outside 
the band obtained according to the standard criteria applied. 

One of our most important results was the demonstration of the fact that the ``Borel stability criterion'' (based 
on a self-evident statement that the physical observables should not depend on the Borel parameter --- 
an auxiliary parameter of the method) is a necessary but by far not sufficient criterion to guarantee a 
good extraction of the hadron parameters. Moreover, combined with the assumption of a 
Borel-parameter independent effective continuum threshold it leads to the extraction of very 
inaccurate values of the bound-state parameters. 

The results reported in \cite{lms_2ptsr,lms_3ptsr,m_lcsr,lms_scalar} gave us a solid ground to claim that also in QCD 
the actual accuracy of the method turns out to be 
much worse than the accuracy expected on the basis of the standard criteria. 

Notice, however, that these results contained not only the cautious messages concerning the application 
of the method of sum rules to hadron properties, but also the identification of the main origin of the 
difficulties of the method: these difficulties arise from an 
over-simplified Ansatz for the hadron continuum which is modeled as a perturbative contribution above 
a Borel-parameter 
independent effective continuum threshold. We introduced the notion of the {\it exact\/} 
effective continuum threshold, 
which corresponds to the true bound-state parameters: in a HO model the true hadron parameters --- 
decay constant and form factor --- 
are known and thus the exact effective continuum thresholds for different correlators may be calculated. 
We have demonstrated that 
the exact effective continuum threshold 
(i) is not a universal quantity and depends on the correlator considered (i.e., it is in general 
different for two-point and three-point vacuum-to-vacuum correlators, and for vacuum-to-hadron correlators), 
and (ii) depends on the Borel parameter and, in the case of the form factor, also on the momentum transfer. 

The understanding of this fact allowed us to make a step forward and to propose a new algorithm for 
extracting ground-state parameters from correlators in those cases where the ground-state energy is 
known\footnote{In our opinion the knowledge of the ground-state mass is mandatory for a successful 
application of the method of sum rules to bound states. Therefore any attempt to study the existence 
of the ground state with the method of sum rules does not yield trustable results.}: 
a simple idea of \cite{lms_prl} is to relax the standard assumption of a Borel-parameter 
independent effective continuum threshold, 
and allow for a Borel-parameter dependent quantity. The parameters of the $T$-dependent 
effective continuum threshold then may be found by fitting the average energy of the dual correlator 
(i.e. the correlator containing only the contribution of the low-energy region below the effective 
continuum threshold) 
to the known energy of the bound state. 
In \cite{lms_prl,lms_npa} we have shown that the application of this idea for vacuum-to-vacuum correlators 
leads to a considerable increase of the actual accuracy of the method. 

In the present paper we extend this idea to the extraction 
of hadron form factors from vacuum-to-hadron correlators \cite{stern}. These correlators are the basic objects for 
the method of light-cone sum rules, which has been extensively applied in the recent years to the analysis 
of $B$-decay form factors \cite{braun,ball}. The issue of the accuracy of light-cone sum rules in QCD is extremely 
important, since the results from this method are extensively used in precision electroweak physics \cite{cern}.

We demonstrate that the application of our algorithm yields two crucial improvements: 
first, it increases significantly the accuracy of the extracted form factors; 
second, it enlarges considerably the range of the momentum transfers where the form factor 
may be extracted from vacuum-to-hadron correlators. 

%\vspace{-.5cm}
\section{The model}
Let us recall the essential formulas: 
We consider a non-relativistic HO model defined by the Hamiltonian ($r\equiv|\vec r\,|$)
\begin{eqnarray}
H = H_0 + V(r), \quad H_0 = {\vec p}^{\,2}/2m, \quad V(r) = {m \omega^2 r^2}/2,
\end{eqnarray}
where all characteristics of the bound states are calculable. 
For instance, for the ground (g) state one finds 
\begin{eqnarray}
\label{EG} 
E_{\rm g} = \frac{3}{2} \omega
,\quad 
\psi_{\rm g}(r) = \left( \frac{m \omega}{\pi} \right)^{3/4} \exp(-m\omega^2 r^2/2),\quad 
F_{\rm g}(q) = \exp(-q^2 / 4m \omega),
\end{eqnarray}
where the elastic form factor of the ground state is defined according to ($q\equiv |\vec q|$)
\begin{eqnarray}
F_{\rm g}(q) = \langle \psi_{\rm g}|J(\vec q)|\psi_{\rm g} \rangle = \int d^3k\,
\psi_{\rm g}^\dagger(\vec k) ~ \psi_{\rm g}(\vec k - \vec q), 
\end{eqnarray}
with the current operator $J(\vec q)$ given by the kernel
\begin{eqnarray}
\label{J} 
\langle \vec r\,'|J(\vec q)|\vec r\rangle = \exp(i\vec q \cdot \vec r) ~ 
\delta^{(3)}(\vec r - \vec r\,').
\end{eqnarray}
Notice that the non-relativistic elastic current is conserved, which leads to the relation $F_{\rm g}(q=0)=1$. 

%*********************************************************************

\section{The vacuum-to-hadron correlator}
In order to apply the method of sum rules to hadron form factors at intermediate and large monentum transfers 
a Borelized vacuum-to-hadron amplitude of the $T$-product of two quark currents is used.  
The analogue of this quantity in quantum mechanics has the form \cite{m_lcsr}
%\begin{eqnarray}
%A(E,q)=\langle \vec r=0|G(E)J(\vec q)|\Psi_{\rm g}\rangle, 
%\end{eqnarray}
%or, after Borelization ($E\to T$), 
\begin{eqnarray}
\label{A}
A(T,q)=\langle \vec r=0|G(T)J(\vec q)|\psi_{\rm g}\rangle. 
\end{eqnarray}
where $G(T)\equiv \exp(-HT)$ is the evolution operator in imaginary time $T$ (i.e., the Borel transform 
$E\to T$ of the full 
Green function of the model, $(H-E)^{-1}$) and $J$ is the current operator defined by (\ref{J}).

An obvious disadvantage of this correlator compared with vacuum-to-vacuum correlators is the necessity 
to know for its calculation the 
ground-state wave function. As a bonus, this correlator receives an enhanced contribution of the ground state, 
which makes it potentially more attractive for the extraction of the ground-state form factor. 
   
The spectral representation for $A(T,q)$ may be written in the form 
\begin{eqnarray}
\label{disp}
A(T,q)=\int d^3k \;G(\vec k^2,T)\psi_{\rm g}(\vec k-\vec q)=
\int_0^\infty dz\,a(z,T,q),\quad z\equiv {\vec k}^2/2m, 
\end{eqnarray}
where \cite{nsvz}
\begin{eqnarray}
\label{G}
G(\vec k^2,T)\equiv \langle \vec r=0|G(T)|\vec k\rangle=
\frac{1}{(2\pi)^{3/2}}\frac{1}{[\cosh(\omega T)]^{3/2}}\exp\left(-\frac{\vec k^2}{2 m \omega}\tanh(\omega T)\right). 
\end{eqnarray}
Explicit expressions for $a(z,T,q)$ and $A(T,q)$ may then be easily 
obtained using $\psi_{\rm g}$ of Eq.~(\ref{EG}). For $A(T,q)$ one finds 
\begin{eqnarray}
\label{Aexact}
A(T,q)=\left(\frac{m\omega}{\pi}\right)^{3/4}\exp\left(-\frac{3}{2}\omega T\right)
\exp\left(-\frac{q^2}{4m\omega}\left(1-e^{-2\omega T}\right)\right). 
\end{eqnarray}
The function $A(T,q)$ depends on two dimensionless variables ${\hat q}^2\equiv {q^2}/{m\omega}$ and $\omega T$. 

Expanding in Eq.~(\ref{disp}) the Green function $G(\vec k^2,T)$ in powers of $\omega$ generates the analogue 
of the QCD twist expansion for the amplitude $A(T,q)$ (see \cite{m_lcsr} for details).

The ground-state contribution to the correlator has the form  
\begin{eqnarray}
A_{\rm g}(T,q)=\psi_{\rm g}(r=0)\exp\left({-E_{\rm g} T}\right)F_{\rm g}(q), 
\end{eqnarray}
so one finds 
\begin{eqnarray}
\label{Rground}
\frac{A_{\rm g}(T,q)}{A(T,q)}=\exp\left(-\frac{q^2}{4m\omega}e^{-2\omega T}\right). 
\end{eqnarray}
Notice the following features of the correlator $A(T,q)$: 

\noindent
(i) At large $T$, the ground state provides the dominant contribution to $A$, similar to the case of 
vacuum-to-vacuum correlators:  
\begin{eqnarray}
A(T,q)\to A_{\rm g}(T,q), \quad \mbox{for} \quad T\to \infty. 
\end{eqnarray}
(ii)
At small $q$, because of current conservation, the ground state dominates the correlator for all $T$. 
This is a specific feature of $A$ which arises due to the choice of the initial state. 
Thus, compared with vacuum-to-vacuum correlators, the correlator $A$ ``maximizes'' the ground-state contribution.  

%%%%%%%%%%%%%%%%%%%%%%%%%%%%%%%%%%%%%%%%%%%%%%%%%%%%%%%%%%%%%%%%%%%%%%%%%%%%%%%%%%%
\section{Extraction of the ground-state form factor}

Employing the quark--hadron duality hypothesis, which assumes that the excited-state 
contribution is dual to the high-energy region of quark diagrams, one constructs the 
correlator dual to the ground state and gets the following sum rule for the form factor $F_{\rm g}(q)$:\footnote{
This is the standard but not the unique way to define the dual correlator: another possibility  --- 
to apply the cut only to the leading-twist contribution and thus to attribute the contributions of all higher 
twists exclusively to the ground state --- has been discussed in \cite{m_lcsr}.}
\begin{eqnarray}
\label{sr}
A^{\rm dual}(T,q,z_{\rm eff})\equiv\int_{0}^{z_{\rm eff}(T,q)} a(z,T,q)dz
=\psi_{\rm g}(r=0)\exp\left(-{E_{\rm g}}T\right)F_{\rm g}(q).
\end{eqnarray}
Relation (\ref{sr}) constitutes the definition of the exact effective continuum thresholds $z_{\rm eff}(T, q)$.

Let us emphasize the following point: As is clear from Eq.~(\ref{disp}), the $T$-dependence of the spectral 
density $a(z,T,q)$ is due to the $T$-dependence of the evolution operator $G(z,T)$ of Eq.~(\ref{G}); 
the latter may be expanded in a series of exponential functions with increasing powers. 
So, the integration over $z$ of $a(z,T,q)$ in Eq.~(\ref{sr}) can produce a single exponential corresponding 
to the ground state only if the effective continuum threshold $z_{\rm eff}$ has a nontrivial $T$-dependence. 

More generally and independently of the specific form of the interaction potential, the 
effective continuum threshold for the vacuum-to-hadron correlator should depend both on $T$ and $q$ as  
a consequence of the structure of the correlator. This may be seen in two ways: first, 
the dominance of the ground state in the correlator at small $q$ implies 
$z_{\rm eff}(T,q\to 0)\to\infty$ for all $T$. Second, for large $T$, the analytic properties of the dual correlator 
(which is a hand-made nonperturbative object whose properties are rather different from 
properties of correlators in perturbation theory \cite{lms_npa}) 
require that $z_{\rm eff}(T\to\infty, q)\to\infty$ for all $q$. 

\begin{figure}[t]
\begin{tabular}{cc}
\includegraphics[width=8cm]{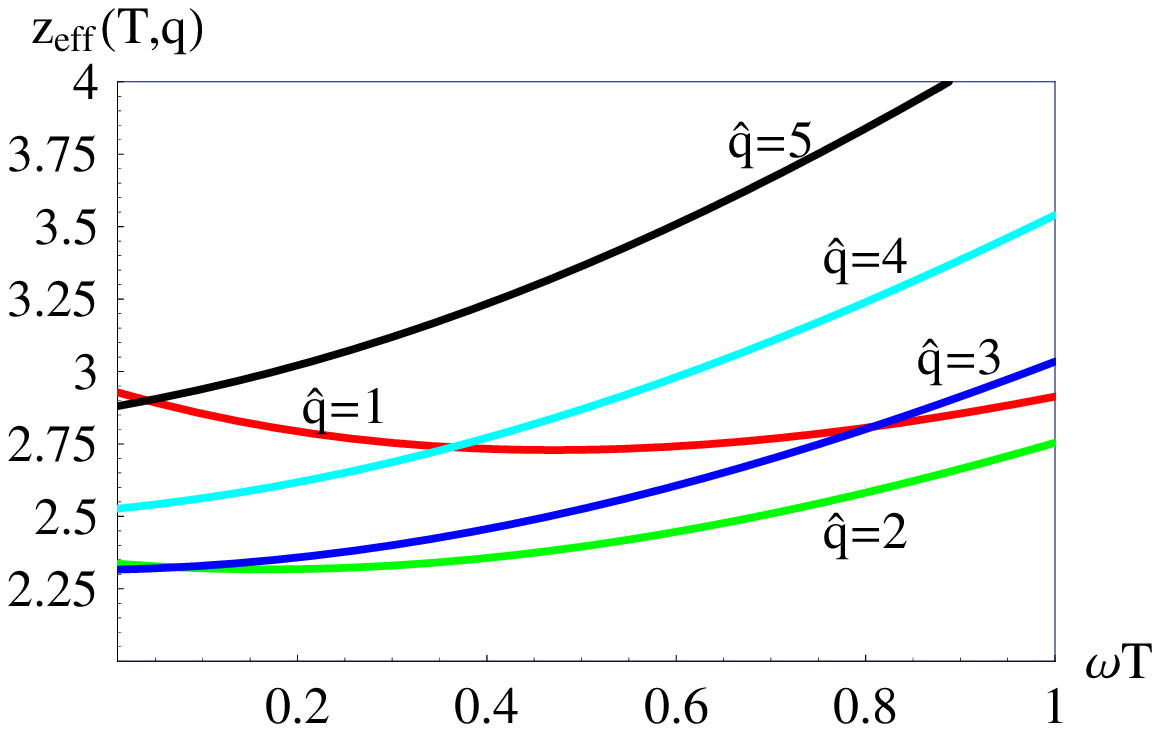} 
&
\includegraphics[width=8cm]{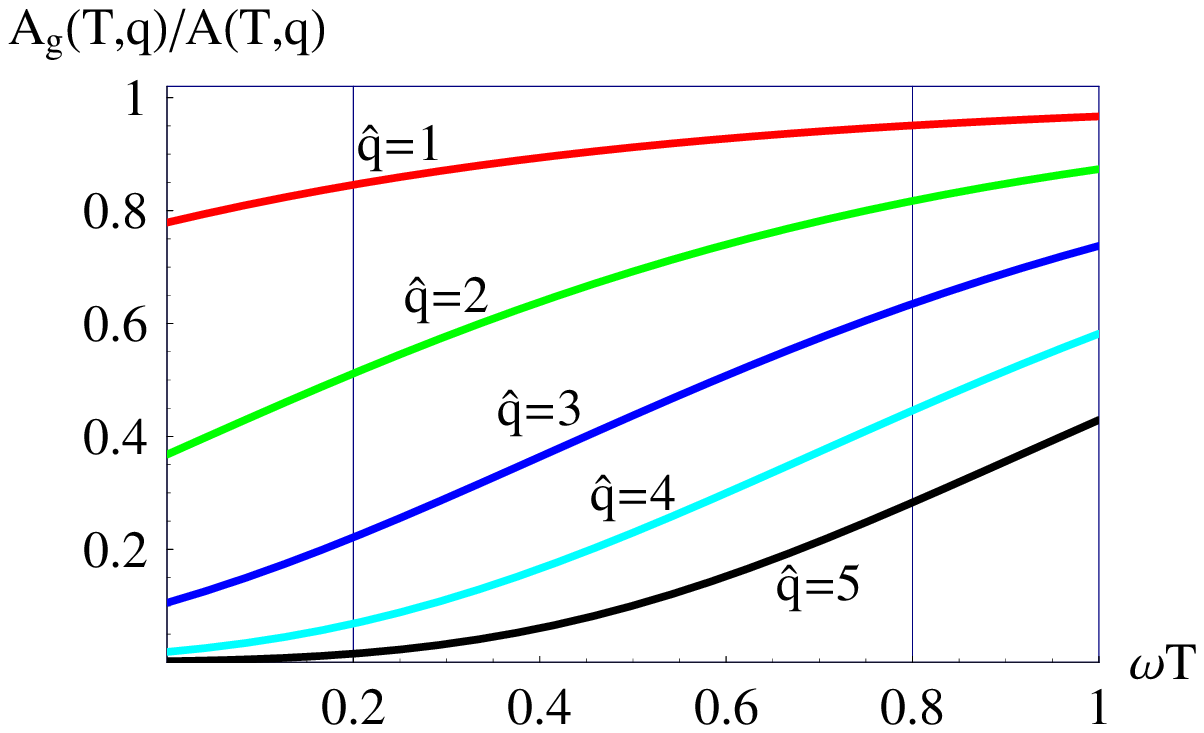} \\
(a) & (b)
\end{tabular}
\caption{\label{fig:zeff}
(a) Exact effective continuum threshold $z_{\rm eff}(T, q)$ obtained 
by solving numerically Eq.~(\ref{sr}) using the exact bound-state parameters 
$\psi_{\rm g}(r=0)$ and $F_{\rm g}(q)$ vs.\ Euclidean time $T$ for several values of $\hat q\equiv q/\sqrt{m\omega}$.
(b) Relative contribution of the ground state to the correlator $A_{\rm g}(T,q)/A(T,q)$ vs.\ $T$ for 
several values of $q$. The vertical lines indicate the window $0.2\le  \omega T \le 0.8$.
Red: $\hat q=1$, 
green: $\hat q=2$, 
blue: $\hat q=3$, 
light-blue: $\hat q=4$, 
black: $\hat q=5$.}
\end{figure}

The explicit $T$- and $q$-dependences of $z_{\rm eff}(T,q)$ can be obtained by solving Eq.~(\ref{sr}) 
making use of the exact bound-state parameters $\psi_{\rm g}(r=0)$ and $F_{\rm g}(q)$. In the HO model, this 
can be easily done numerically. 
The corresponding results are shown in Fig.~\ref{fig:zeff}(a). Obviously, the effective continuum 
threshold $z_{\rm eff}(T, q)$ does depend strongly on both $T$ and $q$. 

Let us consider now a restricted problem where the energy $E_{\rm g}$ of the ground state and its wave 
function at the origin, $\psi_{\rm g}(r=0)$, are known, and try to determine its elastic form factor 
from the sum rule (\ref{sr}).

First, according to \cite{svz} we should determine the working interval of $T$ --- the Borel ``window'' 
where the method may be applied to the extraction of the ground-state parameter: 

\noindent
(i) The upper boundary of the $T$-window is obtained from the condition that the 
truncated expansion gives a good approximation to the exact correlator. 
Since in the HO model the correlator is known exactly, the upper boundary 
is $T = \infty$. 
However, to be close to realistic situations, where only a limited number of higher-twist corrections is available, 
following our results reported in \cite{m_lcsr} we define the upper boundary of the window as 
$\omega T \lesssim 0.8$. 

\noindent
(ii) the lower boundary of this $T$-window is determined by the condition that the 
ground state gives a ``sizable'' contribution to the correlator.\footnote{Notice, however, 
that in applications of light-cone sum rules in QCD the magnitude of the ground-state contribution 
is usually not considered.} As can be seen from Fig.~\ref{fig:zeff}(b), 
if we require the ground-state contribution to exceed, say, 50\%, the window disappears already 
for $\hat q = q/\sqrt{m\omega} > 4$. 
%Thus, requiring a substantial ground-state contribution to the correlator 
%does not allow one to study the region of large momentum transfers. 
We shall see, however, that our algorithm allows one to extract the form factor, 
although with a worse accuracy, also in the region where the ground-state contribution to the correlator 
is below 30\%. Thus, our algorithm opens the possibility to study also the region of large momentum transfers.  

For practical purposes we need to set a lower boundary of the working region, so we choose the window to be 
$0.2 \lesssim \omega T \lesssim 0.8$. 

\begin{figure}[t]
\begin{tabular}{cc}
\includegraphics[width=8cm]{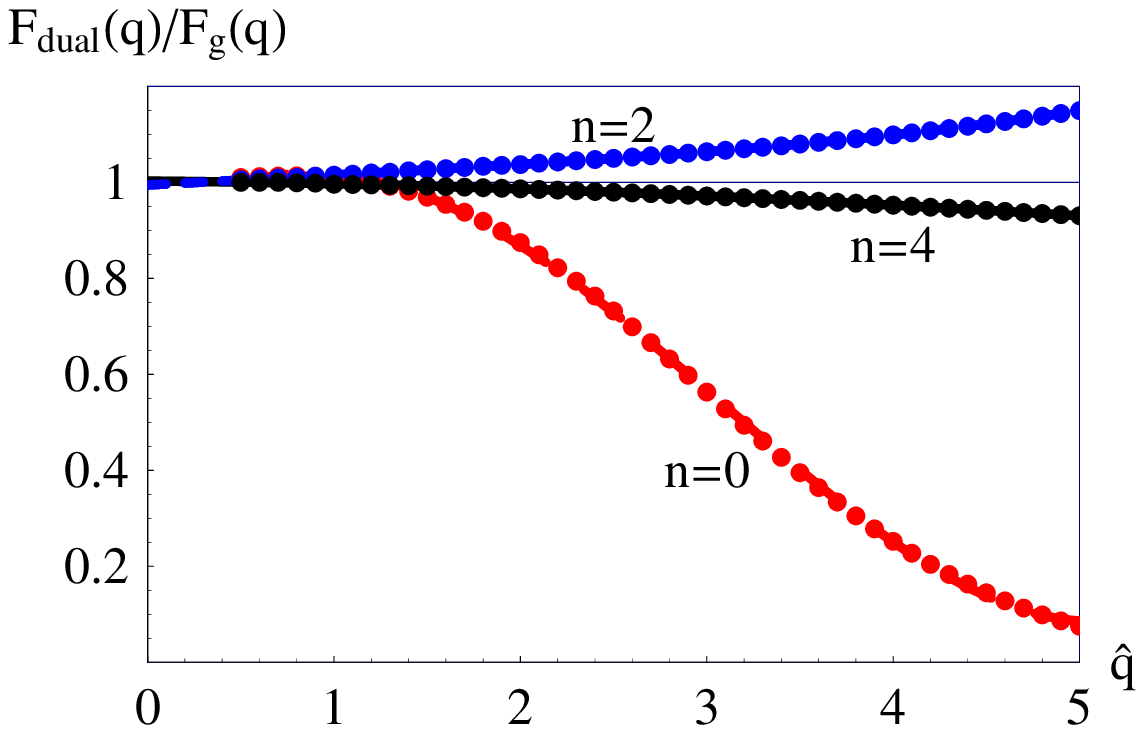}
& 
\includegraphics[width=8cm]{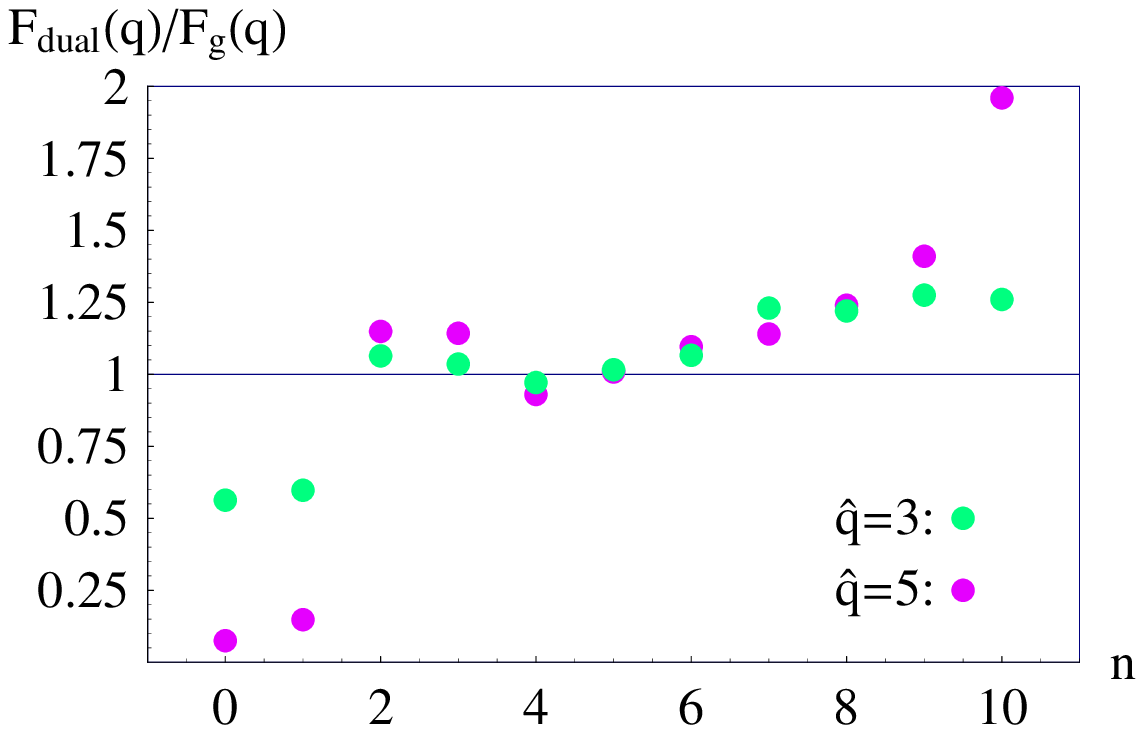}\\
(a) & (b)
\end{tabular}
\caption{\label{fig:FF}
(a) Ratio $F_{\rm dual}(q)/F_{\rm g}(q)$, extracted from the sum rule (\ref{sr}) 
using different approximations $z^{(n)}_{\rm eff}(T, q)$ in Eq.~(\ref{zeff}), vs. $\hat q$. 
Different lines correspond to different Ans\"atze for the effective continuum threshold: 
red --- constant ($n=0$), 
blue --- quadratic ($n=2$), black --- quartic ($n=4$).
(b) Ratio $F_{\rm dual}(q)/F_{\rm g}(q)$, extracted from the sum rule (\ref{sr}) 
using different approximations $z^{(n)}_{\rm eff}(T, q)$ in Eq.~(\ref{zeff})
depending on $n$, the power of the polynomial used in the fit.
The results for two values of the momentum transfer are presented: 
$\hat q=3$ --- green dots, $\hat q=5$ --- violet dots.}
\end{figure}

Second, we must choose a criterion to fix the effective continuum threshold $z_{\rm eff}(T, q)$. 
We proceed in the following way: We consider a set of $T$-dependent Ans\"atze for the effective continuum threshold: 
\begin{eqnarray}
\label{zeff}
z^{(n)}_{\rm eff}(T, q)= \sum\limits_{j=0}^{n}z_j^{(n)}(q)(\omega T)^{j}.
\end{eqnarray}
(The standard procedure adopted in all sum-rule applications in QCD is to assume 
a $T$-independent quantity.) Now, at each value of $q$ we fix the parameters on the 
r.h.s of (\ref{zeff}) as follows: we calculate the dual energy 
\begin{eqnarray}
\label{Edual}
E_{\rm dual}(T, q) =  - \frac{d}{d T} \log A_{\rm dual}(T, q, z_{\rm eff}(T, q)),
\end{eqnarray}
for the $T$-dependent $z_{\rm eff}$ of Eq.~(\ref{zeff}). 
Then we evaluate $E_{\rm dual}(T, q)$ at several values of $T = T_i$ ($i = 1,\dots, N$, 
where $N$ can be taken arbitrary large) chosen uniformly in the window. Finally, 
we minimize the squared difference between $E_{\rm dual}$ and 
the exact value $E_{\rm g}$:
\begin{eqnarray}
\label{chisq}
\chi^2 \equiv \frac{1}{N} \sum_{i = 1}^{N} \left[ E_{\rm dual}(T_i, q) - E_{\rm g}\right]^2.
\end{eqnarray}
The results for the dual form factor obtained after optimizing in this way the 
parameters of $z_{\rm eff}$ are shown in Fig.~\ref{fig:FF}. 

\subsection{$T$-independent effective continuum threshold}

Let us consider first the case of the standard $T$-independent approximation.
One obtains the form factor in the region $\hat q\le 1.5$ with better than 10\% accuracy, 
whereas the accuracy falls down rather fast with increasing $q$ (see Fig.~\ref{fig:FF}(a)). 
The real problem is, however, that the magnitude of the error cannot be guessed 
on the basis of the standard criteria adopted in the method: e.g., at $\hat q=2$, the variation of the extracted form 
factor in the window is only about 2\%, which mimics an accurate extraction of the form factor, whereas 
the actual error comprises 15\%. Thus we conclude that, similar to the case of vacuum-to-vacuum correlators, 
the error of the form factor extracted from the vacuum-to-hadron correlator 
cannot be determined from the variation of the dual form factor in the window. 

\subsection{$T$-dependent effective continuum threshold}

We are going to discuss now the quadratic and quartic Ans\"atze, $z^{(2)}_{\rm eff}(T, q)$ and $z^{(4)}_{\rm eff}(T, q)$  
(the results obtained for $n=1$ are close to the results for $n=0$, 
and the results for $n=3$ are close to those for $n=2$).
As can be seen from Fig.~\ref{fig:FF}(a), the form factor may be extracted with reasonable accuracy 
(better than 15\%) up to $\hat q=5$. This relatively high accuracy is reached in spite of the facts that 
(i) the relative contribution of the ground state to the correlator falls down considerably below 50\% 
in the $T$-window, and 
(ii) the form factor at $\hat q=5$ falls down compared to its value at $\hat q=0$ by almost three orders 
of magnitude. 

It should be mentioned that both for $n=2$ and $n=4$ the dual energy $E_{\rm dual}(T,q)$ 
and the dual form factor $F_{\rm dual}(T,q)$ are extremely stable (better than $0.1 \%$) in the window. 
In spite of this stability, the actual error of the extracted form factor for both $n=2$ and $n=4$ is found to be
at the level of 10--15\%. Again, this error could not be guessed on the basis of the standard Borel 
stability criterion, which thus does not work for vacuum-to-hadron correlators. 
Previously, we made a similar observation for the case of vacuum-to-vacuum 
correlators \cite{lms_2ptsr,lms_3ptsr,lms_prl}.
Interestingly, the {\it actual\/} error may be probed if one compares the results obtained for quadratic and 
quartic Ans\"atze for $z_{\rm eff}$; moreover, in the case under consideration 
these results provide the {\it actual band of values} which contains the true hadron form factor. 
%Therefore the application of this algorithm to form factors within the method of light-cone sum rules in QCD 
%seems very promising. 

Fig.~\ref{fig:FF}(b) shows the results for the form factor obtained by the fit as function of the power $n$ of the 
$T$-polynomial in (\ref{zeff}) for two values $\hat q=3$ and $\hat q=5$. Let us notice that the fitting 
procedure becomes unstable for $n\ge 7$: the fit leads to coefficients 
$z^{(n)}_j$ in (\ref{zeff}) which start to increase with $j$. 
Respectively, increasing the power of the 
polynomial does not lead to an improvement of the accuracy of the extracted from factor. 
This behaviour might be expected 
and reflects the fact that the extraction of the form factor from the correlator 
known in the limited range of $T$ cannot be done arbitrarily accurate. 
%%%%%%%%%%%%%%%%%%%%%%%%%%%%%%%%%%%%%%%%%%%%%%%%%%%%%%%%%%%%%%%%%%%%%%%%%%%%%%%%%%%%%%%%%%%%%%%%%%%%%%
%\newpage
\section{Conclusions}
We discussed the extraction of the form factor from the vacuum-to-bound-state correlator, 
making use of a quantum-mechanical potential model. We formulated a new algorithm for fixing 
the effective continuum threshold --- one of the basic ingredients of the method of sum rules --- 
which to a great extent determines the numerical value of the form factor obtained by this method. 

The main messages of our analysis are as follows: 

$\bullet$
We would like to emphasize that the exact effective continuum threshold in vacuum-to-bound-state correlators, 
(as well as the exact continuum threshold in vacuum-to-vacuum correlators) depends on the Borel parameter $T$. 
This conclusion %is absolutely solid since it 
is based on both the general properties of the vacuum-to-hadron correlator and 
on explicit calculations. 

$\bullet$
Assuming a $T$-independent (i.e., a Borel-parameter independent) Ansatz for the effective 
continuum threshold, the error of the ground-state form factor extracted 
from the vacuum-to-hadron correlator turns out to be typically much larger than 
(i) the error of the description of the exact correlator by the truncated twist expansion and 
(ii) the variation of the dual form factor in the Borel window. 
Combining this statement with our previous results obtained for vacuum-to-vacuum correlators 
\cite{lms_2ptsr,lms_3ptsr,lms_prl,lms_npa}, we conclude 
that in all versions of sum rules ``Borel stability'' cannot be used as a criterion of the reliability 
of the extracted form factor. 

%\item 

$\bullet$
Allowing for a $T$-dependent effective continuum threshold and fixing it according to Eq.~(\ref{chisq}) 
leads to dramatic improvements in the extracted ground-state form factor: 

First, it extends considerably the range of the momentum transfers where the form factor may be 
extracted with an accuracy at the level of 10--15\%. In particular, it extends the applicability of the method 
also to the region of large momentum transfers, where the ground state does not give the dominant contribution 
to the correlator. 

Second, it improves considerably the accuracy and reliability of the form-factor extraction compared 
to the standard $T$-independent 
Ansatz for the effective continuum threshold that has been used in all previous applications of sum rules in QCD. 
Moreover, in the example under consideration, we managed to obtain the band of values which contains the true 
form factor, by analysing different Ans\"atze for the effective continuum threshold. 
If this property holds in QCD, it would eventually provide a means to gain a rigorous control 
over the accuracy of light-cone sum rules. This important issue requires further investigation. 

The application of the proposed ideas to vacuum-to-hadron correlators in QCD seems very promising. 
We believe that it will lead to a considerable improvement of hadron form factors obtained 
from light-cone sum rules. 

%\vspace{.5cm}

%\noindent

\acknowledgments
D.~M.\ gratefully acknowledges financial support from the Austrian Science Fund (FWF) under project P20573. 
The work was supported in part by EU Contract No. MRTN-CT-2006-035482, ``FLAVIAnet''.


\newpage
\begin{thebibliography}{30}
\bibitem{svz}
M.~Shifman, A.~Vainshtein, and V.~Zakharov, Nucl.~Phys.~{\bf B147}, 385 (1979).
\bibitem{ioffe}
B.~L.~Ioffe and A.~V.~Smilga, Phys.~Lett.~{\bf B114}, 353 (1982). \\
V.~A.~Nesterenko and A.~V.~Radyushkin, Phys.~Lett. {\bf B115}, 410 (1982).
\bibitem{ck}P.~Colangelo and A.~Khodjamirian, {\it QCD sum rules: a modern perspective}, hep-ph/0010175.
%\bibitem{iamin}
%M.~Jamin and B.~O.~Lange, Phys.~Rev.~{\bf D65}, 056005 (2002). 
%\bibitem{ball}
%P.~Ball and R.~Zwicky, Phys.~Rev.~{\bf D71}, 014015 (2005).
\bibitem{nsvz}V.~Novikov, M.~Shifman, A.~Vainshtein, and
V.~Zakharov, Nucl.~Phys.~{\bf B237}, 525 (1984).
\bibitem{nsvz1}
A.~I.~Vainshtein, V.~I.~Zakharov, V.~A.~Novikov, and M.~A.~Shifman, 
Sov.~J.~Nucl.~Phys. {\bf 32}, 840 (1980).  
\bibitem{qmsr}
V.~A.~Novikov {\it et al.}, Phys.~Rep.~{\bf 41}, 1 (1978); 
M.~B.~Voloshin, Nucl.~Phys.~{\bf B154}, 365 (1979); 
%On dynamics of heavy quarks in nonperturbative QCD vacuum
J.~S.~Bell and R.~Bertlmann, 
Nucl.~Phys.~{\bf B177}, 218 (1981); 
Nucl.~Phys.~{\bf B187}, 285 (1981);
V.~A.~Novikov, M.~A.~Shifman, A.~I.~Vainshtein, V.~I.~Zakharov, 
Nucl.~Phys.~{\bf B191}, 301 (1981). 
% Are all hadrons alike?
\bibitem{orsay}
A.~Le Yaouanc {\it et al.}, 
Phys.~Rev.~{\bf D62}, 074007 (2000); 
Phys.~Lett.~{\bf B488}, 153 (2000); 
Phys.~Lett.~{\bf B517}, 135 (2001).
\bibitem{ms_inclusive}
D.~Melikhov and S.~Simula , Phys.~Rev.~{\bf D62}, 074012 (2000). 
\bibitem{radyushkin2001}A.~V.~Radyushkin, %Introduction to QCD sum rule approach, 
in {\it Strong Interactions at Low and Intermediate Energies}, edited by J.~L.~Goity, 
Singapore, World Scientific, pp. 91--150 (2000) [hep-ph/0101227].
\bibitem{bakulev}A.~P.~Bakulev, Acta Phys.~Polon.~{\bf B37}, 3603 (2006) [hep-ph/0610266]. 
%************************
\bibitem{lms_2ptsr}
W.~Lucha, D.~Melikhov, and S.~Simula, 
Phys.~Rev.~{\bf D76}, 036002 (2007); 
Phys.~Lett.~{\bf B657}, 148 (2007); \\
Phys.~Atom.~Nucl.~{\bf 71}, 1461 (2008).
\bibitem{lms_3ptsr}
W.~Lucha, D.~Melikhov, and S.~Simula, 
Phys.~Lett.~{\bf B671}, 445 (2009).
\bibitem{m_lcsr}
D.~Melikhov, Phys.~Lett.~{\bf B671}, 450 (2009).
\bibitem{lms_scalar}
W.~Lucha, D.~Melikhov, and S.~Simula, 
Phys.~Rev.~{\bf D75}, 096002 (2007); 
Phys.~Atom.~Nucl.~{\bf 71}, 545 (2008).\\
W.~Lucha and D.~Melikhov, 
Phys.~Rev.~{\bf D73}, 054009 (2006); 
Phys.~Atom.~Nucl.~{\bf 70}, 891 (2007). 
\bibitem{lms_prl}
W.~Lucha, D.~Melikhov, and S.~Simula,  
Phys.~Rev.~{\bf D79}, 096011 (2009). 
\bibitem{lms_npa}
W.~Lucha, D.~Melikhov, and S.~Simula, arXiv:0905.0963 [hep-ph]. 
\bibitem{stern}
N.~S.~Craigie and J.~Stern, Nucl.~Phys.~{\bf B216}, 209 (1983).
%
\bibitem{braun}V.~M.~Braun, A.~Khodjamirian, and M.~Maul, 
Phys.~Rev.~{\bf D61}, 073004 (2000). 
%\bibitem{jamin}M.~Jamin and B.~Lange, Phys.~Rev.~{\bf D65}, 056005 (2002).
\bibitem{ball}P.~Ball and R.~Zwicky, Phys.~Rev.~{\bf D71}, 014015 (2005).
\bibitem{cern} 
M.~Artuso {\it et al.}, Eur.~Phys.~J.~{\bf C57}, 309 (2008). 
\end{thebibliography}
\end{document}